\documentclass[global,twocolumn]{svjour}
\usepackage{graphics}
\journalname{Applied Physics B}
\begin{document}
\title{Classical Simulation of Relativistic Quantum Mechanics in Periodic Optical Structures}
\author{Stefano Longhi
}                     
%
%
\institute{Diaprtimento di Fisica, Politecnico di Milano, Piazza L.
da Vinci 32, I-20133 Milano (Italy)}
%
%
\maketitle
\begin{abstract}
Spatial and/or temporal propagation of light waves in periodic
optical structures offers a rather unique possibility to realize in
a purely classical setting the optical analogues of a wide variety
of quantum phenomena rooted in relativistic wave equations. In this
work a brief overview of a few optical analogues of relativistic
quantum phenomena, based on either spatial light transport in
engineered photonic lattices or on temporal pulse propagation in
Bragg grating structures, is presented. Examples include spatial and
temporal photonic analogues of the Zitterbewegung of a relativistic
electron, Klein tunneling, vacuum decay and pair-production, the
Dirac oscillator, the relativistic Kronig-Penney model, and optical
realizations of non-Hermitian extensions of relativistic wave
equations.
\end{abstract}
\section{Introduction}
\label{sec:1} Quantum-classical analogies have been explored on many
occasions to mimic at a macroscopic level many quantum phenomena
which are currently inaccessible in microscopic quantum systems
\cite{Dragomanbook}. The study of analogies, besides to be fruitful
in gaining insights into different phenomena in nature, provides a
noteworthy strategy in research capable of transferring ideas,
concepts and techniques among apparently unrelated physical fields.
In particular, in the past two decades engineered photonic
structures have provided a useful laboratory tool to investigate and
visualize with classical optics the dynamical aspects embodied in a
wide variety of coherent quantum phenomena encountered in atomic,
molecular, condensed-matter and matter-wave physics
\cite{Longhi09LPR}. The study of quantum-optical analogies has been
greatly stimulated by the development of the field of discrete
optics, aimed to realize novel functionalized optical materials
based on evanescently-coupled optical waveguides in which the
dispersion and diffraction properties of light can be specifically
managed \cite{B01,L1,Szameit10}. The development of recent and
reliable technologies in waveguide fabrication, notably the one
based on waveguide inscription in transparent glasses based on
femtosecond laser writing \cite{Szameit10}, has enabled to access in
the lab to a wide variety of fascinating optical analogues of
quantum systems. Such analogies have also transferred into optics
some important ideas and methods for molding the flow of light
originally developed in the context of quantum control
\cite{Longhi09LPR,Longhibook}. Among the wide variety of
quantum-optical analogies investigated in the last two decades, we
mention the optical analogues of electronic Bloch oscillations
\cite{B01,B02} and Zener tunneling \cite{ZT,Dreisow09}, dynamic
localization \cite{DL}, coherent enhancement and destruction of
tunneling \cite{CDT}, adiabatic stabilization of atoms in
ultrastrong laser fields \cite{stabi}, Anderson localization
\cite{AL}, quantum Zeno effect \cite{QZ}, Rabi flopping \cite{Rabi},
coherent population transfer \cite{STIRAP}, coherent vibrational
dynamics \cite{molecule}, geometric potentials \cite{geometric}, and
dynamical (Kapitza) trapping \cite{Kapitza}. Most of such analogies
are based on the formal similarity between the paraxial optical wave
equation in dielectric media and the single-particle nonrelativistic
Schr\"{o}dinger equation, thus providing a test bed for classical
analogue studies of nonrelativistic quantum phenomena. Optical
analogues of many-body phenomena in nonrelativistic models, such as
photonic analogues of the famous Bose-Hubbard model, have been
proposed as well \cite{Longhi11}. \par

Recently, a great attention has been devoted toward the
investigation of experimentally-accessible and controllable
classical or quantum systems that simulate certain fundamental
phenomena rooted in the relativistic wave equations, such as the
Dirac equation for fermionic particles. Among others, cold trapped
atoms, ions and graphene have proven to provide accessible systems
to simulate relativistic physics in the lab, and a vast literature
on this subject has appeared in the past few years (see, for
instance, \cite{GR1,GR2,ion,ion2} and references therein). In
particular, low-energy nonrelativistic two-dimensional electrons in
graphene obey the Dirac-Weyl equation and behave like massless
relativistic particles. This has lead to the predictions in
condensed-matter physics and atom optics of phenomena analogous to
Zitterbewegung \cite{Zitterbewegung} and Klein tunneling
\cite{Klein} of relativistic massive or massless particles, with the
first experimental evidences of Klein tunneling in graphene
\cite{EGR1}, carbon nanotubes \cite{EGR2}, trapped ions
\cite{KleinIons}, and of Zitterbewegung \cite{ZBion}. \par

 Photonic analogues of Dirac-type equations have been also theoretically proposed
for light propagation in certain triangular or honeycomb photonic
crystals, which mimic conical singularity of energy bands of
graphene \cite{Haldane,Zhang08,Segev09}, as well as in metamaterials
\cite{meta}, optical superlattices
\cite{Longhi09un,Longhi10Klein,Longhi10Pair,DreisowPRL2010}, Bragg
gratings \cite{BR1,BR2,BR3}, and nonlinear quadratic media
\cite{LonghiJPB2010}. Such studies have motivated extended
investigations on the properties of 'photonic graphene'
\cite{Haldane,Segev09,photonicgraphene} and to the proposals of
photonic analogues of relativistic phenomena like Zitterbewegung
\cite{Zhang08,meta,Longhi09un,LonghiJPB2010}, Klein tunneling
\cite{Segev09,meta,Longhi10Klein,BR1}, decay of the quantum vacuum
and pair production \cite{Longhi10Pair}, the Dirac oscillator
\cite{BR2}, and the relativistic versions of the Kronig-Penney model
and surface Tamm states \cite{BR3}. Noticeably, the introduction of
gain and/or loss regions in the optical medium can be exploited to
realize in a classical setting certain non-Hermitian relativistic
models proposed in the context of non-Hermitian quantum mechanics
and quantum field theories \cite{Longhi2010PRL,SzameitTachions}.
\par
 In this article a brief overview of a few important optical analogues of relativistic quantum
phenomena is presented. Among the various optical realizations of
relativistic wave equations proposed in the recent literature and
briefly mentioned in previous discussion, the present work focuses
mostly on a few rather simple periodic optical structures, namely
spatial light transport in evanescently-coupled waveguide arrays and
temporal pulse propagation in Bragg grating structures. Specific
examples of quantum-optical analogies discussed in the work include
spatial and temporal photonic analogues of Zitterbewegung, Klein
tunneling, instability of the quantum vacuum and pair-production,
the Dirac oscillator, the relativistic Kronig-Penney model, and
optical simulations of non-Hermitian relativistic wave equations.

\section{Photonic Zitterbewegung}
Originally predicted by Schr\"{o}dinger in the study of the Dirac
equation \cite{ZBS}, Zitterbewegung (ZB) refers to the trembling
motion of a freely moving relativistic quantum particle that arises
from the interference between the positive (electron) and negative
(positron)  energy states of the spinor wavefunction
\cite{Zitterbewegung,Grenier}. For a free electron, the Dirac
equation predicts the ZB to have an extremely small amplitude (of
the order of the Compton wavelength $\simeq 10^{-12} \; {\rm m}$)
and an extremely high frequency ($\simeq 10^{21} \;{\rm Hz}$),
making such an effect experimentally inaccessible. Moreover, the
physical relevance of ZB in relativistic quantum mechanics is a
rather controversial issue, because such an effect arises in the
framework of the single-particle picture of the Dirac equation, but
not in quantum field theory. Phenomena analogous to ZB, which
underlie the same mathematical model of the Dirac equation, have so
far predicted in a wide variety of quantum and even classical
physical systems, as discussed in the introduction. Here we briefly
present two optical simulations of the relativistic ZB, based the
former on spatial light propagation in binary waveguide arrays, the
latter on the frequency conversion process of optical pulses in
quadratic nonlinear media.

\subsection{Photonic Zitterbewegung in binary waveguide arrays}
The simplest example in optics where one can find an analogue of the
relativistic ZB is may be the discrete transport of light waves in a
one-dimensional binary array \cite{Longhi09un}. Such an optical
system behaves, in fact, like a one-dimensional superlattice, in
which the two minibands of the superlattice plays the same role as
the positive- and negative-energy branches of the Dirac equation
\cite{Cannata90}. Under excitation by a broad light beam near the
Bragg angle, the discretized light propagates along the array
showing a characteristic trembling motion, which mimics the
relativitsic ZB of the electronic wave function. Such an oscillatory
motion has been recently observed in Ref.\cite{DreisowPRL2010} in
binary arrays written in fused silica by the femtosecond laser
writing technology \cite{Szameit10}, providing the first simulation
in optics of ZB.\par Let us consider light propagation in a binary
waveguide array, realized by two interleaved lattices A and B, as
shown in Fig.1(a). In practice, the superlattice can be realized by
a sequence of equally-spaced waveguides with alternating
deep/shallow peak refractive index changes $dn_1$ and $dn_2$. In the
tight-binding approximation, light transport is described by the
coupled-mode equations \cite{Longhi09un,DreisowPRL2010}
\begin{equation}
i \frac{dc_l}{dz}=-\kappa (c_{l+1}+c_{l-1})+(-1)^l \sigma c_l,
\end{equation}
where $c_l$ are the modal field amplitudes in the various
waveguides, $2 \sigma$ and $\kappa$ are the propagation constant
mismatch and the coupling rate between two adjacent waveguides of
the array, respectively. The superlattice supports two minibands,
separated by a narrow gap of width $2 \sigma$ [see Fig.1(b)],
defined by the dispersion curves $\omega_{\pm}(q)= \pm
\sqrt{\sigma^2+4 \kappa^2 \cos^2 (qa)}$, where $2a$ is the lattice
period and $q$ the Bloch wave number. Hence, in the vicinity of the
edges of the first Brillouin zone, e.g. near $q=\pi/(2a)$, the
dispersion curves of the two minibands form two opposite hyperbolas,
and therefore mimic the typical hyperbolic energy-momentum
dispersion relation for positive-energy and negative-energy branches
of a freely-moving relativistic massive particle (dotted graph).
This suggests that light transport in the lattice for Bloch waves
with wave number $q$ close to $\pi/(2a)$ simulates the temporal
dynamics of the relativistic Dirac equation. When launching a tilted
broad beam $E(x,z)$ at the Bragg angle
$\theta_B\simeq\lambda/(4n_sa)$ ($n_s$ is the substrate refractive
index) into the array only a small region around $q=\pi/(2a)$ in
$q$-space is excited. After setting $c_{2n}(z)=(-1)^n\psi_1(n,z)$
and $c_{2n-1}=-i(-1)^n \psi_2(n,z)$ and introducing the continuous
transverse coordinate $\xi\leftrightarrow n=x/(2a)$, the
two-component spinor $\psi(\xi,z)=(\psi_1,\psi_2)^T$ satisfies the
one-dimensional Dirac equation \cite{Longhi09un,Cannata90}
\begin{equation}
i \frac{\partial \psi}{\partial z}=-i \kappa \sigma_x \frac{\partial
\psi}{\partial \xi}+\sigma \sigma_z \psi \;,
\end{equation}
where $\sigma_x=\left(
\begin{array}{cc}
0 & 1 \\
1 & 0
\end{array}
\right)$ and $ \sigma_z=\left(
\begin{array}{cc}
1 & 0 \\
0 & -1
\end{array}
\right)$ are the $\sigma_x$ and $\sigma_z$ Pauli matrices. Equation
(2) corresponds to the one-dimensional Dirac equation for a
relativistic freely-moving particle of mass $m$ \cite{Grenier}
provided that the formal change
\begin{equation}
\kappa \rightarrow c \; , \; \;  \sigma \rightarrow m c^2 / \hbar,
\end{equation}
is made, and $\xi$ and $z$ are interpreted as the spatial and the
temporal variables, respectively. Therefore, in the optical
superlattice the {\it temporal} evolution of the Dirac spinor wave
function $\psi$ is mapped onto the {\it spatial} evolution along $z$
of the field amplitudes $\psi_1$ and $\psi_2$, describing the
occupation amplitudes of light in the two sub-lattices A and B of
the binary array. Correspondingly, ZB is observed as a quivering
spatial oscillatory motion of the beam center of mass $\langle n
\rangle(z)=\sum_n n|c_n|^2/\sum_n|c_n|^2$. Note that the measurable
quantity $\langle n \rangle(z)$ is directly related to the
expectation value of position for the relativistic particle,
$\langle \xi \rangle (z)=\left( \int d \xi \; \xi
(|\psi_1|^2+|\psi_2|^2) \right) / \left( \int d \xi
(|\psi_1|^2+|\psi_2|^2) \right)$, by the simple relation $ \langle n
\rangle \simeq 2 \langle \xi \rangle +1/2$ \cite{Longhi09un}. Let us
indicate by $E(x,0)$ the electric field envelope that excites the
superlattice at the $z=0$ input. For beam incidence close to the
Bragg angle $\theta_B$, one can write $E(x,0)=G(x) \exp(2 \pi i x
n_s \theta_B / \lambda)$, where the envelope $G$ varies slowly over
the waveguide spacing $a$.  Under such an assumption, an exact
expression for $\langle \xi \rangle (z)$ can be derived and reads
\cite{Longhi09un}
\begin{equation}
\langle \xi \rangle(z) =  \langle \xi \rangle(0)+ v_0 z + 2 \pi
\kappa \sigma^2 \int dk (1/ \epsilon^3) \sin (2 \epsilon z )
|\hat{G}(k)|^2
\end{equation}
where $k=2qa-\pi$ is the shifted transverse momentum,
$\epsilon(k)=\sqrt{\sigma^2+\kappa^2 k^2}$ defines the
energy-momentum dispersion relation of the free relativistic
particle, $\hat{G}(k)=(1/2 \pi) \int d \xi G(2 a \xi) \exp(-ik \xi)$
is the angular spectrum of the beam envelope (normalized such that
$\int dk |\hat{G}(k)|^2=1/4 \pi$), and $v_0=4 \pi \kappa^3 \int dk
(k / \epsilon)^2 |\hat{G}(k)|^2$ is the mean particle speed. The
last oscillatory term on the right hand side of Eq.(4), superimposed
to the straight trajectory defined by the first two terms, is the
ZB. For $\hat{G}(k)$ centered at $k=0$, at leading order Eq.(4)
yields $\langle \xi \rangle(z) \simeq \langle \xi \rangle(0)+v_0
z+(\kappa/2 \sigma) \sin(2 \sigma z)$, i.e. the amplitude and
frequency of ZB are  given by
\begin{eqnarray} \label{ZBAmplitude}
 R_{ZB}=&\;\kappa/(2\sigma)=\hbar/(2mc)\\ \label{ZBFrequency}
 \omega_{ZB}=&2 \sigma=2 m c^2 / \hbar,
\end{eqnarray}
respectively. Therefore, ZB vanishes for either the far-relativistic
($m \rightarrow 0$) and the weak-relativistic ($m \rightarrow
\infty$) limits: in the first case the amplitude of ZB diverges, but
the oscillation frequency $\omega_{ZB}$ goes to zero, whereas in the
latter case the frequency of ZB diverges but its amplitude vanishes
(see, for instance, \cite{Cannata90}). As an example, Figs.1(c) and
(d) show the ZB as obtained by numerical simulations of the
coupled-mode equations for two values of the detuning $\sigma$ and
for a broad Gaussian beam that excites the array at the Bragg angle.
The damping of the oscillations observed in the figures is related
to the spectral angular broadening of the incident beam
\cite{Longhi09un}. The previous analysis can be extended to a
photonic superlattice in two transverse spatial dimensions $x$ and
$y$. The superlattice is composed by two interleaved triangular
lattices, with a propagation constant detuning $\sigma$ between the
guided modes, and a broad input beam is launched at the Bragg angle.
In this case, in the continuous limit the evolution of the beam
envelope along the longitudinal direction $z$ is described by a
two-dimensional Dirac-type equation, which reduces to the massless
case of the previously-studied photonic graphene
\cite{Haldane,Zhang08,Segev09} in the $\sigma \rightarrow 0$ limit.

\begin{figure}
\resizebox{0.45\textwidth}{!}{%
  \includegraphics{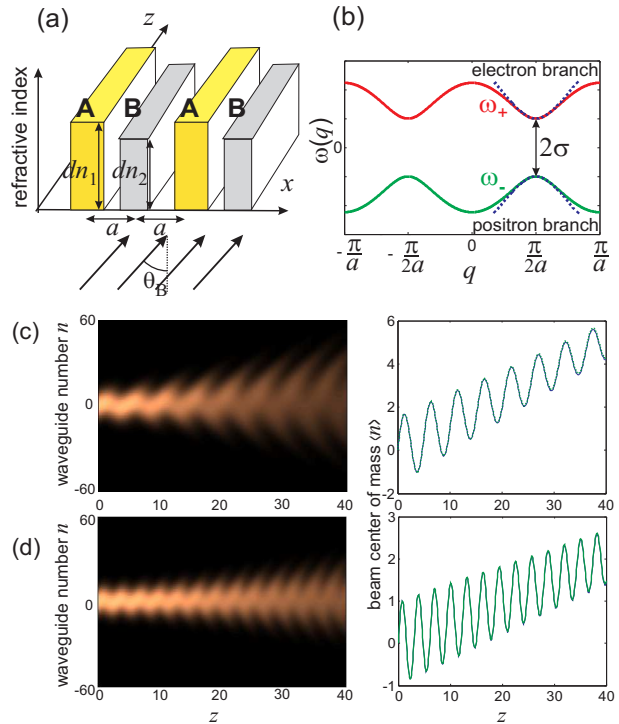}
} \caption{Photonic analogue of relativistic Zitterbewegung in a
binary waveguide array. (a) Schematic of the optical superlattice
[refractive index change profile $n(x)-n_s$]. (b) Dispersion curves
of the two minibands of the superlattice, corresponding to the
electron and positron energy branches of the Dirac equation. (c)
Evolution of a broad Gaussian beam (snapshot of $|c_n(z)|^2$, left
panel), exciting the binary array and tilted at the Bragg angle, and
corresponding beam trajectory (right panel), for parameter values
$\kappa=1$ and $\sigma=0.6$. (d) Same as (c), but for $\sigma=1$.}
\end{figure}

\subsection{Photonic Zitterbewegung in nonlinear frequency conversion}
A second example of a photonic analogue of ZB is provided by the
process frequency conversion of optical pulses in a nonlinear
quadratic medium arising from group velocity mismatch
\cite{LonghiJPB2010}. Let us consider the propagation of three
optical pulses at carrier frequencies $\omega_1$, $\omega_2$ and
$\omega_3=\omega_1+\omega_2$ in a nonlinear quadratic medium and in
presence of group velocity mismatch. To study the analogue of ZB in
the frequency conversion process, we assume that at the input plane
$z=0$ the nonlinear crystal is excited by a strong and nearly
continuous-wave pump field at frequency $\omega_1$, and by a weak
and short signal pulse at frequency $\omega_2$ and temporal profile
$g(t)$. Assuming that group velocity dispersion is negligible and
assuming perfect phase matching, in the no-pump-depletion
approximation the sum-frequency generation process in described by
the two coupled wave equations
\begin{eqnarray}
\left( \frac{\partial}{\partial z}+\frac{1}{v_{g2}}
\frac{\partial}{\partial t} \right) A_2& = & -i \kappa A_3 \\
\left( \frac{\partial}{\partial z}+\frac{1}{v_{g3}}
\frac{\partial}{\partial t} \right) A_3 & = & -i \kappa A_2
\end{eqnarray}
where $A_l$ ($l=2,3$) is the amplitude of the electric field
envelope at frequency $\omega_l$, normalized such that $|A_l|^2$
($l=2,3$) is the photon flux at frequency $\omega_l$, $v_{gl}$ is
the group velocity in the medium at frequency $\omega_l$, $\kappa
=\rho {\sqrt {I_1/ \hbar \omega_1}}$, $\rho$ is the strength of the
nonlinear interaction, and $I_1$ is the intensity of the pump field.
The coupled equations (7) and (8) can be cast in a Dirac form after
introduction of the coordinates of a moving frame
\begin{equation}
\xi=z \; , \; \; \eta=t-z/v_g
\end{equation}
where the velocity $v_g$ is defined by the relation
\begin{eqnarray}
\frac{1}{v_g}=\frac{1}{2} \left(
\frac{1}{v_{g2}}+\frac{1}{v_{g3}}\right).
\end{eqnarray}
In the moving frame, after introduction of the spinor wave field
$\psi=(A_2,A_3)^T$ Eqs.(7) and (8) can be cast in the form of a
Dirac equation
\begin{equation}
i \frac{\partial \psi}{\partial \xi}=-i \sigma_z \delta
\frac{\partial \psi}{\partial \eta}+\kappa \sigma_x \psi
\end{equation}
where $\sigma_x$ and $\sigma_z$ are the Pauli matrices and where we
have set
\begin{equation}
\delta=\frac{1}{2} \left( \frac{1}{v_{g2}}-\frac{1}{v_{g3}}\right).
\end{equation}
Note that, after the formal change
\begin{eqnarray}
\delta & \rightarrow  & c \nonumber \\
\kappa & \rightarrow  & \frac{m c^2}{\hbar} \\
 \xi & \rightarrow & t \; ,\;  \eta  \rightarrow  x \nonumber
\end{eqnarray}
Eq.(11) corresponds to the one-dimensional Dirac equation for a
relativistic particle of mass $m$ in absence of external fields,
moving alon the $x$ axis, written in the Weyl representation
\cite{Grenier}. Such a representation is different, though
equivalent, to the one used in the previous subsection [compare
Eq.(11) with Eq.(2)], the spinor components in the two
representations  being related by a unitary transformation.
Therefore, the {\it temporal} evolution of the spinor wave function
$\psi$ for the Dirac particle is mapped into the {\it spatial}
evolution of the envelopes $A_2$ and $A_3$ for signal and
sum-frequency pulses, respectively, whereas the spatial coordinate
of the Dirac particle is mapped into the retarded time $\eta$ of the
optical pulses. In the absence of group-velocity mismatch, the
massless Dirac equation is obtained, in which ZB vanishes. As an
example, Fig.2 shows ZB in the process of sum frequency generation
which applies to a periodically-poled Liyhoum Niobate (PPLN) crystal
with a quasi-phase matched grating that ensures phase matching at
the wavelengths $\lambda_1=1550$ nm (pump), $\lambda_2=810$ nm
(signal), and $\lambda_3=532$ nm (sum-frequency). The crystal is
pumped by a near continuous-wave pump of intensity $I=1 \; {\rm
MW}/{\rm cm}^2$ and excited at the input plane by a short Gaussian
signal pulse of duration $\tau_p=0.5$ ps. The mismatch of the group
velocities of signal and sum-frequency fields clearly results in an
oscillating motion (jitter) of the signal pulse along the
propagation in the nonlinear crystal, which is the signature of ZB.

\begin{figure}
\resizebox{0.48\textwidth}{!}{%
  \includegraphics{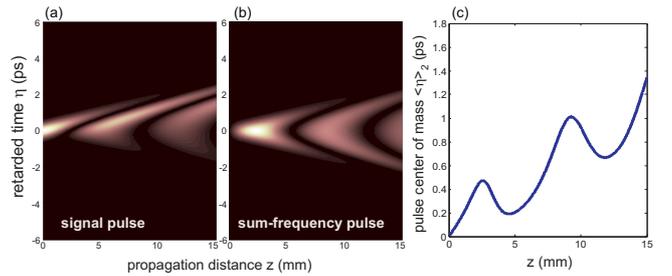}
} \caption{Zitterbewegung in the process of sum frequency generation
induced by group velocity mismatch. (a) and (b) show the evolution
of the signal and sum-frequency pulse intensities in a 15-mm-long
PPLN crystal for a continuous-wave pump. The crystal is excited by a
Gaussian signal pulse of duration $\tau_p=0.5$ ps. In (c) the
evolution of the pulse center of mass $\langle \eta \rangle_1$ for
the signal wave is depicted. The oscillatory behavior of $\langle
\eta \rangle_1$, arising from the group velocity mismatch between
signal and sum-frequency fields, is a signature of ZB.}
\end{figure}

\section{Photonic analogues of Klein tunneling}
A remarkable prediction of the Dirac equation is that a
below-barrier electron can pass a large repulsive and {\em sharp}
potential step without the exponential damping expected for a
non-relativistic particle. Such a transparency effect, originally
predicted by Klein \cite{Klein} and referred to as Klein tunneling
(KT), arises from the existence of negative-energy solutions of the
Dirac equation and requires a potential step height $\Delta V$ of
the order of twice the rest energy $mc^2$ of the electron
\cite{Calogeracos99}. Relativistic tunneling across a {\em smooth}
potential step, which describes the more physical situation of a
constant electric field $E$ in a finite region of space of length
$l$, was subsequently studied by Sauter \cite{Sauter}. Sauter showed
that to observe barrier transparency the potential increase $\Delta
V \simeq eEl$ should occur over a distance $l$ of the order or
smaller than the Compton wavelength $\lambda_C=\hbar/(mc)$, the
transmission probability rapidly decaying toward zero for a smoother
potential increase \cite{Calogeracos99,Sauter,Emilio}. The required
field corresponds to the critical field for $e^+e^-$ pair production
in vacuum, and its value is extremely strong making the observation
of relativistic KT for electrons very challenging. Therefore,
growing efforts have been devoted to find experimentally accessible
systems to investigate analogs of relativistic KT, as discussed in
the introduction section. In particular, great interest has
suscitated the proposal and first experimental evidences of KT for
non-relativistic electrons in graphene, which behave like massless
Dirac fermions. In optics, several proposals of KT analogs have been
suggested as well in the past few years, as discussed in the
introduction section. An important optical set up is provided by
light transport in honeycomb photonic lattices, the co-called
photonic graphene, whose band structure mimics the one of graphene
\cite{Segev09}. However, even in simpler one-dimensional periodic
optical structures one can realize optical analogues of KT
\cite{Longhi10Klein,BR1}. Here we briefly discuss two of such simple
realizations. The former one is based on monochromatic light
propagation in waveguide superlattices \cite{Longhi10Klein}, whereas
the latter one is based on temporal pulse propagation in fiber Bragg
gratings \cite{BR1}.

\subsection{Photonic Klein tunneling in optical superlattices}
As in Sec.2.1, let us consider the transport of monochromatic
discretized light waves in a binary array of waveguides, realized by
two interleaved lattices A and B. As compared to the superlattice
used to realize the ZB [Fig.1(a)], here a weak modulation $R_l$ of
the index change, much smaller than either $dn_1$ and $dn_2$ and
equal for the interleaved lattices A and B, i.e. such that
$R_{2l-1}=R_{2l}$, is superimposed to the lattice, as shown in
Fig.3(a). Light transport in the superlattice is then described by
the coupled-mode equations [compare with Eqs.(1)]
\begin{equation}
i \frac{dc_l}{dz}=-\kappa (c_{l+1}+c_{l-1})+(-1)^l \sigma c_l+\Phi_l
c_l,
\end{equation}
\begin{figure}
\resizebox{0.45\textwidth}{!}{%
  \includegraphics{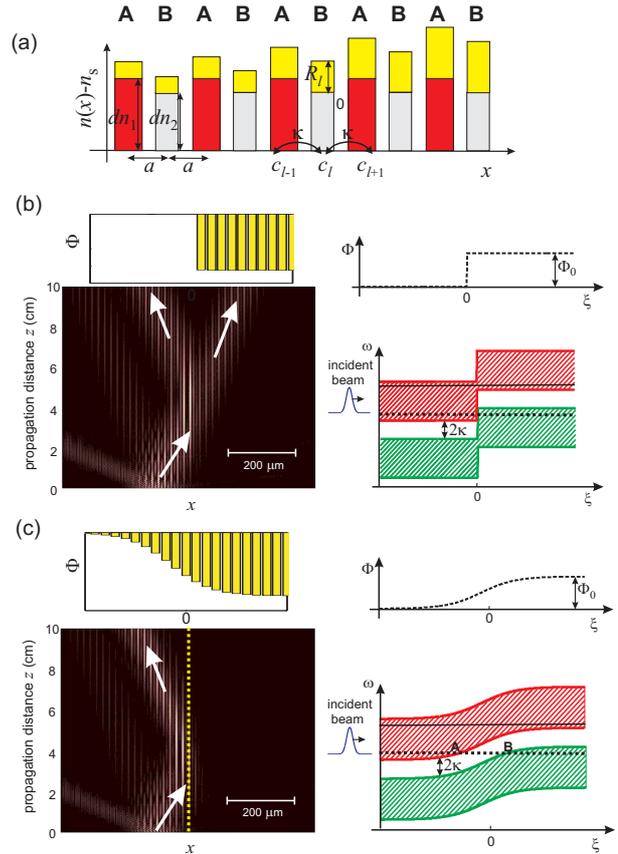}
} \caption{Photonic analogue of relativistic Klein tunneling in a
binary waveguide array with an interface. (a) Schematic of the
optical superlattice. (b) Beam refraction across a potential step,
corresponding to KT. (c) Absence of beam refraction, corresponding
to inhibition of KT, for a smooth potential barrier. The right
panels in (b) and (c) depict the space-energy band diagram, clearly
explaining the inhibition of tunneling in (c).}
\end{figure}
 where the weak modulation $R_l$ of the refractive index change is
responsible, at leading order, to a slight change $\Phi_l$ of the
modal propagation constants  in the various waveguides. For a broad
beam exciting the superlattice tilted close to the Bragg angle, in
the continuous limit light transport turns out to be described by
the Dirac wave equation [compare with Eq.(2)]
\begin{equation}
i \frac{\partial \psi}{\partial z}=-i \kappa \sigma_x \frac{\partial
\psi}{\partial \xi}+\sigma \sigma_z \psi +\Phi(\xi) \psi \;,
\end{equation}
where $\Phi(\xi)=\Phi_{2l} =\Phi_{2l-1}$. Note that Eq.(15) is
formally analogous to the one-dimensional Dirac equation for an
electron of mass $m$ in presence of an electrostatic potential
$\Phi(x)$, once the formal substitutions (3) are made (see, for
instance, \cite{Grenier}). Therefore, by engineering the refractive
index depths of the various waveguides in the array, a step-like
potential $\Phi(\xi)$ at $\xi=0$ can be realized. KT can be observed
as a beam transmission (refraction) across the interface
\cite{Longhi10Klein}. By varying the steepness of the potential
step, the optical superlattice can be designed to demonstrate the
disappearance of KT in a smooth potential step as well, i.e. the
transition from KT for a steep potential to the Sauter's tunneling
suppression for a smooth potential step. In the optical context,
Sauter's inhibition of light tunneling at a smooth interface can be
explained by considering the space-dependent band structures of the
tight-binding superlattices, shown in Fig.3. Such bands are similar
to the energy band diagrams of a semiconductor, in which the two
superlattice minibands (i.e. the electron and positron energy
branches of the Dirac equation) play the role of the conduction and
valence bands and the effect of the external potential $\Phi(\xi)$
is to curve the band structure. For a sharp potential step
[Fig.3(b)] at the interface $\xi=0$ the electron and positron energy
bands are overlapped, and the beam does not need to cross any
forbidden region, i.e. KT occurs. However, owing to the sharp
discontinuity of media properties,  beam transmission is not
complete, and some light is reflected like in a Fresnel
discontinuity between two different dielectric media. Conversely,
for a smooth potential step [see Fig.3(c)] the beam has to cross a
forbidden region, which behaves like a potential barrier. The width
of the potential barrier is indicated by the segment AB in Fig.3(c).
Because the width AB increases as the potential step gets smoother,
the corresponding tunneling probability, i.e. beam transmission
across the interface, rapidly decreases. This explains the
inhibition of KT and the transition to the Sauter's regime
\cite{Sauter}.

\subsection{Photonic Klein tunneling in fiber Bragg gratings}
Another rather simple optical realization of Klein tunneling is
provided by photonic tunneling in structured fiber Bragg gratings
(FBGs) \cite{BR1}. In a FBG, the effective index of the fiber is
modulated along the longitudinal $z'$ direction according to $
n(z')=n_0+\Delta n \; m(z') \cos[2 \pi z' / \Lambda +2 \phi(z')]$,
where $n_0$ is the effective mode index in absence of the grating,
$\Delta n \ll n_0$ is the peak index change of  the grating,
$\Lambda$ is the nominal period of the grating defining the
reference frequency $\omega_B=\pi c/(\Lambda n_0)$ of Bragg
scattering, $c$ is the speed of light in vacuum, and $m(z')$, $2
\phi(z')$ describe the slow variation, as compared to the scale of
$\Lambda$, of normalized amplitude and phase, respectively, of the
index modulation. Note that the local spatial frequency of the
grating is $k(z')=2 \pi / \Lambda+2(d \phi/dz')$, so that the local
chirp rate is $C=dk/dz'=2(d^2 \phi/dz'^2)$. The periodic index
modulation leads to Bragg scattering between two counterpropagating
waves at frequencies close to $\omega_B$. By letting
$E(z',t)=\varphi_1(z',t) \exp[-i \omega_B t +ik_B z'+i \phi(z')]+
\varphi_2(z',t) \exp[-i \omega_B t -ik_B z'-i \phi(z')]+c.c.$ for
the electric field in the fiber, where $k_B=\pi/\Lambda$, the
envelopes $\varphi_1$ and $\varphi_2$ of counterpropagating waves
satisfy coupled-mode equations. After introduction of the
dimensionless variables $z=z'/Z$ and $\tau=t/T$ , with
characteristic spatial and time scales $Z=2 n_0/(k_B \Delta n)$ and
$T=Z/v_g$, and the new envelopes $\psi_{1,2}(z')=[\varphi_1(z') \mp
\varphi_2(z')]/ \sqrt 2$, the coupled-mode equations can be cast in
the Dirac form \cite{BR1}
\begin{equation}
i \partial_{\tau} \psi=-i \sigma_x \partial_z \psi+m(z) \sigma_z
\psi +V(z) \psi
\end{equation}

\begin{figure}
\resizebox{0.45\textwidth}{!}{%
  \includegraphics{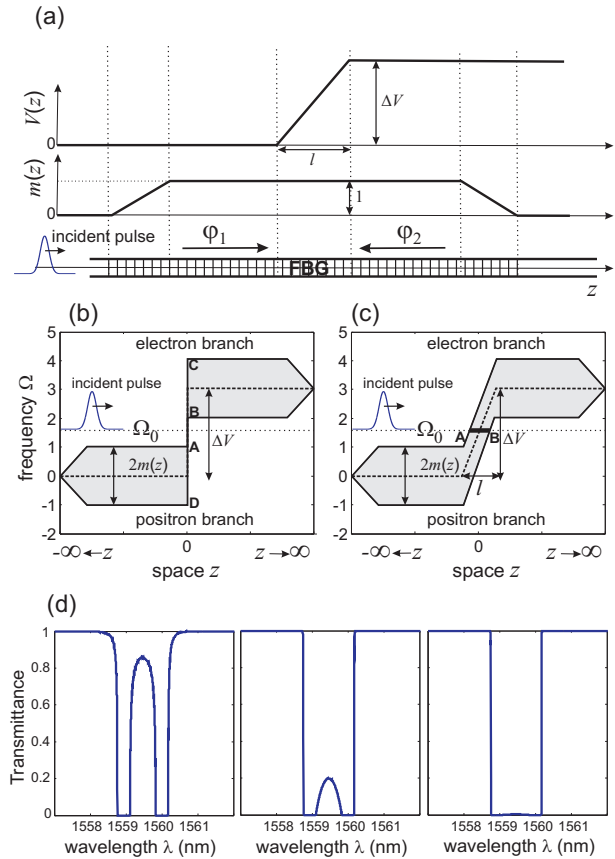}
} \caption{Photonic realization of relativistic Klein tunneling in a
structured FBG. (a) Schematic of the local period detuning $V=(d
\phi/dz)$ (upper plot) and amplitude $m$ (lower plot) profiles of a
FBG to observe the optical analogue of KT. (b) and (c): Space-energy
band diagrams of the FBG for a sharp [(b)], and a smooth [(c)]
potential step $V(z)$ of height $\Delta V$. The shaded regions are
the forbidden (stop band) energies, the dotted horizontal line is
the frequency $\Omega_0$ of the incoming wave packet, and the dashed
curve is the shape of the potential step $V(z)$. (d) Numerically
computed spectral transmittance of three FBGs for increasing length
of the chirped region (from left to right). The transmission window
inside the stop band of the grating for a sharp potential step (left
panel) is the signature of KT.}
\end{figure}

for the spinor wave function $\psi=(\psi_1,\psi_2)^T$, where
$V(z)=(d \phi /dz)$, $v_g$ is the group velocity at the Bragg
frequency, and $\sigma_{x,z}$ are the Pauli matrices. In its present
form, Eq.(16) is formally analogous to the one-dimensional Dirac
equation with $\hbar=c=1$ in presence of an external electrostatic
potential $V(z)$,  $m$ playing the role of a dimensionless (and
generally space-dependent) rest mass (see, for instance,
\cite{Calogeracos99,Emilio}). The optical analogue of the forbidden
energy region of the Dirac equation, in this case, simply
corresponds to the photonic stop band of the periodic grating. As
the refractive index modulation of the grating, i.e. the mass term
$m$ in the Dirac equation (16), is decreased, the stop band region
shrinks and the limit of a massless Dirac equation (similar to the
one describing the dynamics of electrons in graphene near a Dirac
point) is attained. The additional external potential $V$ in
Eq.(16), related to the phase modulation of the grating according to
$V(z)=(d \phi/dz)$, changes the local position of the forbidden
energy region. Therefore, pulse propagation in a FBG with a suitably
designed phase and amplitude grating profiles can be used to mimic
the relativistic tunneling of a wave packet in a potential step
$V(z)$. A typical diagram of a structured FBG suited to realize KT
and the inhibition of KT for a smooth potential step, together with
the space-energy diagram of the grating \cite{BR1}, is shown in
Figs.4(a-c). The signatures of KT and of its inhibition for a smooth
potential step are simply revealed by spectrally-resolved
transmission measurements of the grating. In fact, for a sharp
potential step a transmission window is opened inside the bandgap of
the periodic grating, whereas for a smooth potential step tunneling
is prevented [according to the diagram of Fig.4(c)] and the
transmission window inside the gap vanishes. This is shown in
Fig.4(d), which depicts typical transmission spectra in three FBGs
with different chirp rates, realized to operate near the
$\lambda=1.5 \; \mu$m wavelength of optical communications (for more
details see \cite{BR1}).

\section{Optical simulation of vacuum decay and pair production}

Electron-positron pair production due to the instability of the
quantum electrodynamics (QED) vacuum in an external electric  field
is another remarkable prediction of Dirac theory and regarded as one
of the most intriguing nonlinear phenomena in QED, whose
experimental observation is still lacking (see, for instance,
\cite{libro1,libro2}). In intuitive terms and in the framework of
one-particle Dirac theory, the pair production process can be simply
viewed as the transition of an electron of the Dirac sea occupying a
negative-energy state into a final positive-energy state, leaving a
vacancy (positron) in the negative-energy state. There are basically
two distinct transition mechanisms: the Schwinger mechanism, induced
by an ultrastrong static electric  field, and dynamic pair creation,
induced by time-varying electric fields. The Schwinger mechanism
\cite{Swinger} is basically a tunneling process through a
classically forbidden region, bearing a close connection to Klein
tunneling discussed in the previous section. The other mechanism,
namely dynamic pair creation, was originally proposed by Brezin and
Itzykson \cite{Brezin} for oscillating spatially homogeneous fields,
and has recently attracted great attention because of its potential
implementation using counter-propagating ultrastrong laser fields.
Dynamic pair production is closely related to such intriguing
phenomena like Rabi oscillations of the Dirac sea (see, for
instance, \cite{libro2}). In the framework of the one-particle Dirac
theory of vacuum decay, a simple picture of pair production is
represented by the time evolution of an initially negative-energy
Gaussian wave packet, representing an electron in the Dirac sea,
under the influence of an oscillating electric  field \cite{32}.
When the $e^+e^-$ pair is produced, a droplet is separated from the
wave packet and moves opposite to the initial one \cite{32}. The
droplet is a positive-energy state and represents the created
electron. In a recent work \cite{Longhi10Pair}, it was shown that
light transport in a binary waveguide array with a
sinusoidally-curved optical axis provides a classical simulator of
the dynamic pair production process, in which pair production is
visualized as a breakup of an initial Gaussian wave packet, composed
by a superposition of Bloch modes of the lowest lattice miniband and
representing an electron in the Dirac sea. Periodic axis bending of
the waveguides mimics the effect of an external ac field, which
induces transitions into the upper lattice miniband (the electron
energy branch). The optical structure proposed to simulate dynamic
pair production in basically the binary superlattice of Fig.1(a),
however the optical axis of the waveguides is now periodically bent
along the propagation direction $z$. If the array is excited by a
broad beam tilted close to the Bragg angle, the continuous limit of
the coupled-mode equations now leads to the following Dirac equation
for the spinor $\psi$ \cite{Longhi10Pair} [compare with Eq.(2)]
\begin{equation}
i \frac{\partial \psi}{\partial z}=-i \kappa \sigma_x \frac{\partial
\psi}{\partial \xi}+\sigma \sigma_z \psi \;-2 \kappa \Phi(z)
\sigma_x \psi,
\end{equation}
where $\Phi(z)$ accounts for the waveguide axis bending and reads
explicitly \cite{Longhi10Pair}
\begin{equation}
\Phi(z)=\frac{2 \pi n_s a (dx_0/dz)}{\lambda}.
\end{equation}
In the previous equation, $a$ is the distance between two adjacent
waveguides, $n_s$ is the bulk refractive index, $\lambda$ the
wavelength of light, and $x_0(z)$ is the axis bending profile. Note
that, after the formal change
\begin{eqnarray}
\kappa & \rightarrow & c \nonumber \\
 \sigma & \rightarrow & \frac{ mc^2}{
\hbar} \nonumber \\
\xi & \rightarrow & x \\
\Phi & \rightarrow & \frac{eA_x} {2 \hbar c} \nonumber \\
z & \rightarrow & t, \nonumber
\end{eqnarray}

\begin{figure}
\resizebox{0.45\textwidth}{!}{%
  \includegraphics{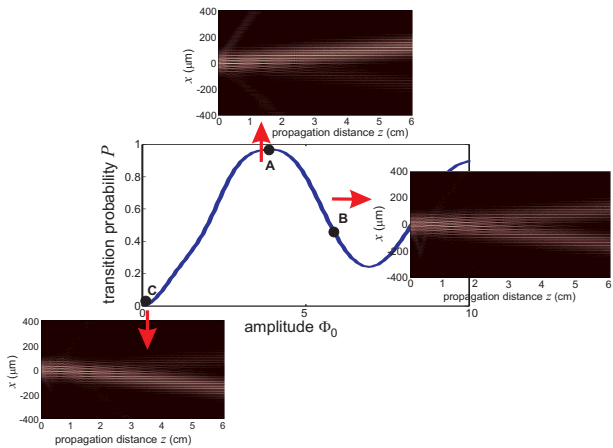}
} \caption{Photonic realization of dynamic $e^+e^-$ pair production
in a binary waveguide array with bent optical axis. The figure shows
the probability of pair production after the application of a
single-cycle ultrastrong laser pulse for increasing values of the
amplitude $\Phi_0$. The insets show the detailed beam evolution
along the array corresponding to the amplitudes of A, B and C. The
fractional beams deflected at opposite angles correspond to states
in the negative (positron) and positive (electron) energy branches
of the Dirac equation.}
\end{figure}
 Eq.(17) corresponds to the one-dimensional Dirac
equation for an electron of mass $m$ and charge $e$ in presence of a
spatially-homogeneous and time-varying vectorial potential
$\mathbf{A}=(A_x,0,0)$, which describes the interaction of the
electron with an external oscillating electric field
$E_x(t)=-(\partial A_x /
\partial t) $ in the dipole approximation (see, for instance,
\cite{Grenier}). Because of momentum conservation, in the
spatially-homogeneous and time-dependent field the problem of pair
creation can be reduced to the transition between two states
consisting of a negative and a positive energy state coupled by the
external field. Within the two-level model, pair production
generally occurs as a multiphoton resonance process enforced by
energy conservation, with interesting effects such as Rabi
oscillations of the quantum vacuum. Pair production induced by
 an ultrastrong and ultrashort laser pulse can be mimicked by
 assuming, for instance, a single-cycle of the ac field [$\Phi(z)=\Phi_0 \sin(2 \pi z/
\Lambda)$ for $0<z<\Lambda$, $\Phi(z)=0$ for $z<0$ and $z>\Lambda$].
As an example, Fig.5 shows the behavior of the transition
probability $P$, after the interaction with a single-cycle pulse, as
obtained by numerical simulations of the Dirac equation for
$\kappa=2$, $\sigma=1.817$, $\Lambda= 0.6676$, momentum $q=\pi/(4a)$
,and for increasing values of the the field amplitude $\Phi_0$. The
detailed beam evolution of an initial Gaussian wave packet along a
waveguide array with single-cycle modulated axis, corresponding to
the three conditions A,B and C of Fig.5, are depicted in the insets
of the figure. The input wave packet is mostly composed by Bloch
modes belonging to the lowest miniband of the array, i.e. it belongs
to the Dirac sea (the negative-energy branch of the Dirac equation).
Owing to the modulation, excitation of the upper moninband,
corresponding to the creation of a $e^+e^-$ pair, is clearly
observed. Note that, as the wave packets belonging to the two
minibands refract at different angles, they separate each other
after some propagation distance. Such a splitting is precisely the
signature of $e^+e^-$ pair production, as discussed in \cite{32}.

\section{Optical simulation of the Dirac oscillator}
 The relativistic extension
of the quantum harmonic oscillator, the so-called Dirac oscillator
(DO) \cite{Dir1,Dir2,Dir3}, provides a paradigmatic and exactly
solvable model in relativistic quantum mechanics. Originally
proposed in quantum chromodynamics in connection to quark
confinement models in mesons and baryons \cite{Dir4}, the DO has
received great interest in relativistic many-body theories and
supersymmetric relativistic quantum mechanics (see
\cite{Dir2,Dir3,Dir5,Dir6,Dir7} and references therein). The DO
model is obtained from the free Dirac equation by the introduction
of the external potential via a non-minimal coupling
\cite{Dir1,Dir3,Dir7}. Since the resulting equation is linear in
both momentum and position operators,  in the nonrelativistic limit
a Schr\"{o}dinger equation with a quadratic potential is then
obtained. In spite of the great amount of theoretical studies, the
DO model in relativistic quantum mechanics and particle physics
remains far from any experimental consideration. In a recent work
\cite{BR2}, a photonic realization of the DO, based on light
propagation in structured FBGs, has been proposed. The main idea is
that, as shown in Sec.3.2, coupled-mode equations describing forward
and backward light coupling in a Bragg grating structure has the
form of a Dirac equation. Using the same notations as in
Ref.\cite{BR2}, let us consider light propagation in a FBG with an
index grating $ n(z)=n_0+\Delta n \; h(z) \cos[2 \pi z / \Lambda +
\phi(z)]$, where $n_0$ is the effective mode index in absence of the
grating, $\Delta n \ll n_0$ is a reference value of the index change
of the grating, $\Lambda$ is the nominal grating period defining the
Bragg frequency $\omega_B=\pi c/(\Lambda n_0)$, $c$ is the speed of
light in vacuum, and $h(z)$, $\phi(z)$ describe the amplitude and
phase profiles, respectively, of the grating. To study Bragg
scattering of counterpropagating waves at frequencies close to
$\omega_B$, let $E(z,t)=\{ u(z,t) \exp(-i \omega_B t +2 \pi i z n_0
/ \lambda_0)+ v(z,t) \exp(-i \omega_B t -2 \pi i z n_0 /
\lambda_0)+c.c. \}$ be the electric field in the fiber, where
$\lambda_0=2 n_0 \Lambda$ is the Bragg wavelength, and the envelopes
$u$ and $v$ of counterpropagating waves satisfy coupled-mode
equations. After introduction of the dimensionless variables $x=z/Z$
and $\tau=t/T$ , with characteristic spatial and time scales
$Z=\lambda_0/(\pi \Delta n)$ and $T=Z/v_g$ ($v_g$ is the group
velocity at the Bragg frequency), and the new envelopes
$\psi_{1,2}(z)=[u(z) \mp v(z)]/ \sqrt 2$, the coupled-mode equations
can be cast in the Dirac form \cite{BR2}
\begin{equation}
i \partial_{\tau} \psi=\sigma_x \left\{ p_x -if(x) \sigma_z \right\}
\psi + \sigma_z m(x) \psi
\end{equation}
for the spinor wave function $\psi=(\psi_1,\psi_2)^T$, where
$\sigma_{x}$ and $\sigma_{z}$ are the Pauli matrices, $p_x=-i
(d/dx)$, and
\begin{equation}
m(x)=h(x) \cos[\phi(x)] \; , \; \; f(x)=-h(x) \sin[\phi(x)].
\end{equation}
In its present form, Eq.(20) is analogous to the one-dimensional
Dirac equation \cite{Grenier}, written in atomic units
($\hbar=c=1$), with a space dependence mass $m$ and with the
momentum operator $p_x$ substituted with $p_x-if(x) \sigma_z$. The
space dependence of the particle mass $m$ is known to describe the
particle interaction with a scalar Lorentz potential, whereas the
substitution $p_x \rightarrow p_x-if(x) \sigma_z$ corresponds to a
non-minimal coupling which is essential to describe the relativistic
DO (see, for instance, \cite{Dir7}). To realize the one-dimensional
analog of the DO, let us choose the amplitude $h$ and phase $\phi$
profiles of the grating such that $h \cos \phi=m_0$ and $h \sin
\phi=-f(x)= \omega_s m_0 x$, i.e. [see Fig.6(b)]
\begin{equation}
h(x)=m_0 \sqrt{1+(\omega_s x)^2} \; , \; \; \phi(x)= {\rm
atan}(\omega_s x),
\end{equation}
where $m_0$ and $\omega_s$ are two arbitrary constants,
corresponding to the particle rest mass and oscillation frequency of
the DO in the non-relativistic limit \cite{Dir7}. Analytical
expressions of the energy spectrum for the one-dimensional DO and of
corresponding bound states can be derived following a standard
procedure detailed e.g. in Ref.\cite{Dir7}. Assuming for the sake of
definiteness $\omega_s>0$, the positive-energy spectrum (electron
branch) of the DO is given by
\begin{equation}
\delta_n= \sqrt{m_0^2+2m_0 \omega_s(1+n)} \; , \; \; n=0,1,2,3,...
\end{equation}
whereas the negative-energy spectrum (positron branch) is given by
\begin{equation}
\delta_n= -\sqrt{m_0^2+2m_0 \omega_s n} \; , \; \; n=0,1,2,3,...
\end{equation}
\begin{figure}
\resizebox{0.45\textwidth}{!}{%
  \includegraphics{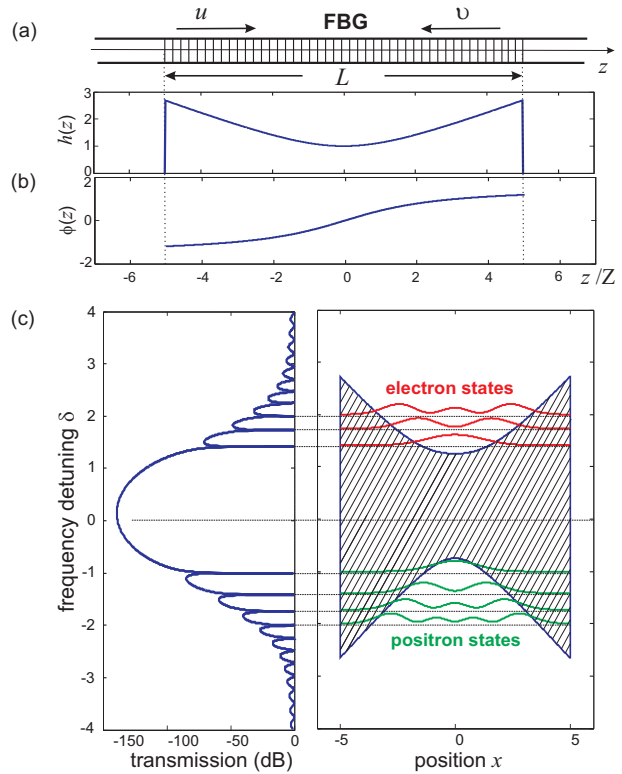}
} \caption{Photonic realization of the Dirac oscillator in a fiber
Bragg grating. (a) Schematic of a FBG. (b) Example of FBG amplitude
and phase profiles that realize the analog of the Dirac oscillator
($m_0=1$, $\omega_s=0.5$, $L/Z=10$). (c) Numerically-computed power
transmission (dB units) of a lossless FBG with amplitude and phase
profiles shown in (b) (left panel), and corresponding reflection
band diagram [dashed area, right panel] with a few low-order
intensity profiles of trapped modes in the electron and positron
branches.}
\end{figure}
The corresponding eigenfunctions $\psi_{\pm}(x)$ can be simply
expressed in terms of Hermite polynomials multiplied by a Gaussian
function. Note that the negative (positron) energy spectrum of the
DO is not obtained from the positive (electron) energy spectrum by
sign reversal ($\delta \rightarrow -\delta)$, the positron branch
possessing an additional bound state with energy $\delta=-m_0$. In
our FBG realization of the DO, bound-states with positive and
negative energies should correspond to trapped light states in the
FBG with resonance frequencies above ($\delta>0$) and below
($\delta<0$) the Bragg frequency $\omega_B$, respectively.  In a FBG
of finite length, the ideal amplitude profile $h(x)$, defined by
Eq.(22), must be truncated, i.e. one has $h(x)=0$ for $|x|>L/(2Z)$,
where $L$ is the grating length [see Fig.6(b)]. The effect of
grating truncation is twofold. First, the bound states of the DO
become actually resonance modes with a finite lifetime, which should
be thus observable as narrow transmission peaks embedded in the stop
band of the grating. Second, the number of resonance modes sustained
by the grating is finite owing to grating truncation. As an example,
Fig.6(c) shows a typical transmission spectrum [power transmission
versus normalized frequency detuning $\delta=(\omega-\omega_B)T$],
for a FBG with length $L/Z=10$ for parameter values $m_0=1$ and
$\omega_s=0.5$. In the figure, the corresponding reflection-band
diagram (dashed area) of the grating in the $(x,\delta)$ plane,
together with the electron ($\delta>0$) and positron ($\delta<0$)
levels and a few eigenstates of the DO, are also depicted. The
transmission peaks visible in the spectrum of Fig.6(c), embedded in
the stop band of the FBG, occur precisely at the values $\delta_n$
predicted by Eqs.(23) and (24). Note the asymmetry of the
transmission spectrum around $\delta=0$, with an additional
resonance in the positron ($\delta<0$) branch of the spectrum with
no counterpart in the electron ($\delta>0$) branch. In an
experiment, the resonant states of the DO and the asymmetry of the
spectrum can be simply detected from spectrally-resolved
transmission measurements of the grating using standard techniques.

\section{Photonic realization of the relativistic Kronig-Penney model and relativistic surface Tamm states}

The Kronig-Penney model for the non-relativistic Schr\"{o}dinger
equation \cite{KP} is one among the simplest models in solid-state
physics that describes the electronic band structure of an idealized
one-dimensional crystal. Relativistic extensions of the
Kronig-Penney model (also referred to as the Dirac-Kronig-Penney
model) have been discussed by several authors (see, for instance,
\cite{RKP,Dr0,Dr1,Dr2,Dr3,Dr4,Dr5,Dr6,Dr7,Dr8,Dr9,Dr10} and
references therein), and the impact of relativity on the band
structure and localization, such as shrinkage of the bulk bands with
increasing band number, have been highlighted on many occasions. In
earlier studies, the Dirac-Kronig-Penney model also attracted some
attention and caused a lively debate about the existence of
so-called Dirac surface states, i.e. relativistic surface Tamm
states which disappear in the non-relativistic limit
\cite{Dr0,Dr4,SS1,SS2,SS3,SS4,SS5,SS6}. In Ref.\cite{BR3}, a
photonic realization of the relativistic Kronig-Penney model and
relativistic surface Tamm states has been proposed, which is based
on light propagation in superstructure FBGs with phase defects.
Light propagation in a FBG with a periodic sequence of phase slips
was shown to simulate the relativistic Kronig-Penney model, the band
structure of which being mapped into the spectral transmission of
the FBG. Similarly, a semi-infinite FBG with phase defects
interfaced with a uniform FBG with a different modulation period was
shown to support Tamm surface states analogous to the relativistic
Tamm states. Such surface states are responsible for narrow
resonance peaks in the transmission spectrum of the grating.\par
Using the same notations as in Sec.3.2, light propagation in a
superstructur FBG is described by the Dirac equation (16), where the
scalar ($m$) and vectorial ($V$) potential terms entering in the
equation are defined by the apodization and phase profiles of the
grating, respectively. The Dirac-Kronig-Penney model for an
infinitely-extended lattice corresponds to a constant mass
$m(z)=m_0$ and to a potential $V(z)$ given by the superposition of
equally-spaced $\delta$-like barriers, namely \cite{RKP}
\begin{equation}
V(z)=V_0 \sum_{n=-\infty}^{\infty} \delta(z-na)
\end{equation}
where $V_0>0$ is the area of the barrier and $a$ is the lattice
period.  Stationary solutions $\psi(z,\tau)=\psi_0(z) \exp(-iE
\tau)$ to the Dirac equation (16) in the periodic potential (25)
with energy $E$ are of Bloch-Floquet type, i.e.
$\psi_0(z+a)=\psi_0(z) \exp(iqa)$, where $q$ is the Bloch wave
number which varies in the first Brilloiun zone ($-\pi/a \leq q <
\pi/a$). The corresponding energy spectrum is composed by a set of
allowed energy bands $E=E(q)$, which are defined by the following
implicit equation (see, for instance, \cite{SS4})
\begin{equation}
\cos(qa)=\cos(V_0) \cos( \kappa a)+\frac{E}{\kappa} \sin(V_0)
\sin(\kappa a),
\end{equation}
where we have set
\begin{equation}
\kappa=\sqrt{E^2-m_0^2}.
\end{equation}
Equation (26) defines the dispersion relation of the relativistic
Kronig-Penney model, which has been investigated by several authors
(see, for instance, \cite{RKP,Dr1,Dr4}). The ordinary
non-relativistic limit of the Kronig-Penney model is attained from
Eqs.(26) and (27) for $V_0 \ll 1$ and for energies $E$ close the
$m_0$, for which the energy-momentum relation (27) reduces to the
non-relativistic one [$E \simeq m_0+\kappa^2/(2m_0)$]; in this
regime, the dispersion relation (26) reduces to
$\cos(qa)=\cos(\kappa a)+(m_0 V_0/\kappa) \sin(\kappa a)$, which is
the ordinary dispersion relation encountered in the non-relativistic
Kronig-Penney model. For larger energies $E$ but still for a low
barrier area $V_0 \ll 1$, non-relativistic effects come into play as
perturbative effects, which modify positions and widths of the
allowed energy bands. Non-relativistic effects deeply modify the
band structure of the crystal for potential strengths $V_0$ of the
order $\sim 1$. In particular, if $V_0$ is an integer multiple of
$\pi$, all band gaps disappear and the dispersion relation reduces
to the one of a relativistic free particle [Eq.(16) with $V(z)=0$],
as if the $\delta$-barriers were absent. In a FBG, the
Dirac-Kronig-Penney model simply corresponds to an
infinitely-extended uniform FBG with a superimposed periodic
sequence of lumped phase slips of equal amplitude $\Delta \phi=2
V_0$ and spaced by the distance $a$. The circumstance that the
effects of the $\delta$ barriers disappear in the
Dirac-Kronig-Penney model when $V_0$ is an integer multiple $\pi$ is
simply due to the fact that, under such a condition, the phase slips
are integer multiplies of $2 \pi$, and thus the grating has no phase
defects and mimics the dynamics of a one-dimensional free
relativistic Dirac particle. A special case of the Dirac Kronig
Penney model corresponds to case $V_0=\pi/2$. In this case, the
resulting superstucture FBG comprises a periodic sequence of $\pi$
phase slips, which has been proposed and demonstrated to realize
slowing down of optical pulses \cite{Longhi1}. The band structure of
the Dirac-Kronig-Penney model is simply mapped into the alternation
of stop/transmission bands observed in spectrally-resolved
transmission measurements of the FBG. An example of the spectral
transmission features of a superstructure FBG with periodic $\pi$
phase slips, in which the allowed spectral transmission bands
reproduce the band structure of the Dirac Kronig Penney model, is
shown in Fig.7. \par

\begin{figure}
\resizebox{0.42\textwidth}{!}{%
  \includegraphics{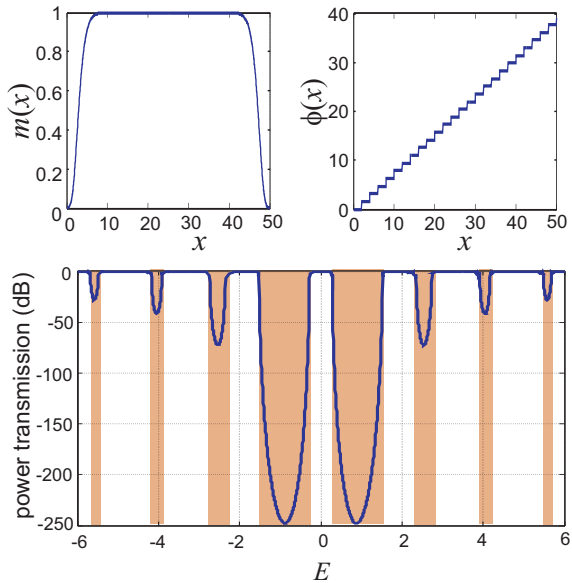}
} \caption{Photonic realization of the Dirac-Kronig-Penney model in
a superstructure FBG comprising a periodic sequence of $\pi$ phase
slips. The figure shows the spectral power transmission of the
grating with amplitude and phase profiles given in the upper insets.
Parameter values are $V_0=\pi/2$, $a=2$, $m_0=1$ and $L=50$. The
dashed areas are the stop bands of the corresponding
Dirac-Kronig-Penney infinite lattice.}
\end{figure}

If the periodic potential $V(z)$ is truncated, surface Tamm states
do appear. Such states for the relativistic Kronig-Penney model
attracted some interest in earlier papers by several authors
\cite{Dr0,Dr4,SS1,SS2,SS3,SS4,SS5,SS6}, and a lively debate was
raised about the proper boundary conditions that should be imposed
to the relativistic wave function at a $\delta$ barrier. As earlier
works \cite{Dr0,SS1,SS2,SS3} suggested that the relativistic
treatment yields a new class of surface states (the so-called Dirac
surface states) which do not correspond the common Tamm states in
the non-relativistic limit, it was subsequently realized that
application of more physical boundary conditions does not yield any
surface state which violates the Tamm condition in the
non-relativistic limit \cite{SS4}. The relativistic extension of the
Tamm model is defined by the potential (see, for instance,
\cite{Dr4,SS4})
\begin{equation}
V(z)= \left \{
\begin{array}{cc}
V_1 & z <0 \\
V_0 \sum_{n=1}^{\infty} \delta(z-na) & z>0
\end{array}
\right.
\end{equation}
Surface states are found as localized solutions to Eq.(16), near the
surface $z=0$, satisfying the appropriate boundary conditions, as
discussed in \cite{SS4}. In our photonic system, the potential
$V(z)$ defined by Eq.(28) and supporting the surface Tamm states at
the $z=0$ boundary is basically realized by two adjacent sections of
uniform grating regions but with different grating periods, with the
second grating region (at $z>0$) comprising a sequence of
equally-spaced phase slips, at a distance $a$, equal to $\Delta
\phi=2V_0$. The existence of surface states can be simply recognized
by the appearance of narrow resonance peaks embedded in a stop band
region of the transmission spectrum of the grating (for more details
we refer to \cite{BR3}).

\section{Photonic realizations of non-Hermitian relativistic wave equations}
Since the the seminal paper by Bender and Boettcher \cite{Bender98},
a great attention has been devoted toward the investigation of
non-Hermitian extensions of quantum mechanics and quantum field
theories. Indeed, many works have remarked that the Hermiticity of
the underlying Hamiltonian can be relaxed, and that a a consistent
quantum theory can be constructed for a broader class of
Hamiltonians
\cite{Bender02,Mostafazadeh02,BenderReview,MostafazadehReview}, in
particular those possessing parity-time ($\mathcal{PT}$) symmetry.
Non-Hermitian Hamiltonians are also very often found in reduced
descriptions of open Hermitian quantum systems, with important
applications to atomic, molecular and condensed-matter physics
\cite{RotterReview}. Several recent works have shown that optical
structures in media with a complex refractive index can provide an
experimentally accessible test bed to simulate in a purely classical
setting non-Hermitian features rooted in
 the non-relativistic Schr\"{o}dinger
 equation with a complex potential \cite{O2,O3,O1,O4,O5,O6,O7}.
Recently, non-Hermitian extensions of relativistic wave equations
\cite{D1,D2,D3} and non-Hermitian quantum field theories
\cite{field} have attracted an increasing interest as well, however
their physical realizations remain mostly unexplored. In
Ref.\cite{Longhi2010PRL}, optical simulations of non-Hermitian wave
equations have been proposed, which are based on light propagation
in distributed-feedback (DFB) optical structures with controlled
gain and/or loss regions. \par Let us indicate by $n(z)=n_0-\Delta n
h(z) \cos(2 \pi z/ \Lambda+2 \theta(z))$ the effective index grating
of the DFB structure, where $n_0$ is the modal refractive index in
absence of the grating, $\Delta n \ll n_0$ and $\Lambda$ are the
peak index change and the nominal period of the grating,
respectively, and $h(z)$, $2 \theta(z)$ are the normalized amplitude
and phase profiles, respectively, of the grating. For a pure index
grating, $h(z)$ is real-valued, whereas for a pure gain grating
$h(z)$ is purely imaginary; in the most general case $h(z)$ can be
taken to be complex-valued \cite{Poladian}. The periodic modulation
of the refractive index leads to Bragg scattering between two
counterpropagating waves at frequencies close to the Bragg frequency
$\omega_B=\pi c/(\Lambda n_0)$. The linear space-dependent
absorption coefficient of counterpropagating waves in the structure
is indicated by $ \alpha_0 (z)$ ($\alpha_0 >0$ in lossy regions,
$\alpha_0<0$ in gain regions). In a semiconductor DFB structure,
gain and loss regions could be tailored by a judicious control of
current injection across the active layer \cite{Poladian}.
Indicating by $E(z,\tau)=\psi_1(z,\tau) \exp[-i \omega_B \tau +ik_B
z+i \theta(z)]+ \psi_2(z,\tau) \exp[-i \omega_B \tau -ik_B z-i
\theta(z)]+c.c.$ the electric field propagating in the DFB
structure, where $k_B=\pi/\Lambda$, the envelopes $\psi_1$ and
$\psi_2$ of counterpropagating waves satisfy coupled-mode equations
\cite{Poladian}. After introduction of the scaled space and time
variables $x=z/Z$ and $t=\tau/T$, with $Z=2n_0 \Lambda/(\pi \Delta
n)$ and $T=Z/v_g$, where $v_g \simeq c/n_0$ is the group velocity of
light at frequency $\omega_B$, the envelopes $\psi_1$ and $\psi_2$
satisfy the following Dirac-type equation in the Weyl representation
\cite{Longhi2010PRL}
\begin{equation}
i \partial_t \psi= -i \sigma_z \partial_x \psi+ \sigma_x m(x)
\psi+V(x) \psi \equiv H \psi
\end{equation}
with complex-valued mass $m$ and vector potential $V$  given by
\begin{equation}
m(x)=h(x) , \; \;  V(x)=\frac{d \theta}{dx}-i \gamma(x),
\end{equation}
where $\gamma(x)=Z \alpha_0 (x)$ is the dimensionless absorption
coefficient and $\sigma_{x,z}$ are the Pauli matrices. In a DFB
structure, both $m(x)$ and $V(x)$ have a limited support over a
spatial length $L$ (the grating region), i.e. $m=V=0$ for $|x|>L/2$.
Since $H$ is non-Hermitian, its spectrum is generally
complex-valued, however it could happen that, in spite of
non-Hermiticity, the energy spectrum remains real-valued, like in
$\mathcal{PT}$ symmetric Hamiltonians in the unbroken symmetry
phase. In addition, because of the non-self-adjointness of $H$,
spectral singularities could arise in the spectrum of $H$, which
correspond to either a zero-width resonance in the
transmission/reflection spectrum of the DFB (the threshold for
self-oscillation) \cite{O5,Longhi2010PRL} or to a perfect absorption
of radiation under a suitable two-port coherent excitation of the
DFB structure \cite{Longhi2010PRL,Longhi2010PRA}. Examples of
non-Hermitian Dirac Hamiltonians showing such a two different kinds
of spectral singularities were discussed in Ref.\cite{Longhi2010PRL}
assuming a pure index grating (i.e. $h(x)$ real-valued) for either
$\mathcal{PT}$-invariant and $\mathcal{PT}$-non-invariant DFB
structures. Another interesting example of a non-Hermitian
relativistic wave equation is obtained by considering a purely
gain-grating ($h$ purely imaginary) in the absence of absorption
losses and chirp ($V=0$). In this case, for an infinitely-long
grating ($L \rightarrow \infty$, $m=im_0$ constant, with $m_0$
real-valued) Eq.(29) represents a superluminal extension of the
relativistic Dirac equation \cite{tachy1} describing a freely-moving
hypothetical tachyonic particle, corresponding to a negative mass
square at rest (see e.g. \cite{tachy2}). A remarkable property of
such an equation, in addition to enable a kind of superluminal
propagation, is the possibility to make a DFB structure fully
transparent, as recently discussed in Ref.\cite{LonghiOL10}.
Tachyonic extensions of Dirac equations can be realized as well in
two-dimensional honeycomb photonic lattices made of coupled
waveguides with alternating gain and loss regions (a complex
photonic graphene), as recently proposed in
Ref.\cite{SzameitTachions}.

\section{Conclusion and outlook}
In this article, a brief overview on the possibility offered by
light transport in periodic photonic structures to simulate in a
purely classical setting the optical analogues of a wide variety of
quantum phenomena rooted in relativistic wave equations has been
presented. Spatial or temporal light transport in engineered
photonic lattices and Bragg grating structures can simulate the
Zitterbewegung of a relativistic electron, Klein tunneling, vacuum
decay and pair-production, the Dirac oscillator, the relativistic
Kronig-Penney model, and certain non-Hermitian extensions of the
Dirac equations, including superluminal (tachyonic) wave equations.
Further quantum-optical analogies are expected to be investigated
and to be implemented in experiments using waveguide lattices and
passive/active Bragg grating and DFB structures.\\
\par
 The author acknowledges financial support by the
Italian MIUR (Grant No. PRIN-2008-YCAAK project "Analogie
ottico-quantistiche in
strutture fotoniche a guida d'onda").\\


\begin{thebibliography}{}

\bibitem{Dragomanbook}
D. Dragoman and M. Dragoman, {\it Quantum-Classical Analogies}
(Springer, Berlin, 2004).

\bibitem{Longhi09LPR}
S. Longhi, Laser \& Photon. Rev. {\bf 3}, 243–261 (2009).

\bibitem{B01}
D. N. Christodoulides, F. Lederer, and Y. Silberberg, Nature {\bf
424}, 817 (2003).

\bibitem{L1}
F. Lederer, G.I. Stegeman, D.N. Christodoulides, G. Assanto M.
Segev, and Y. Silberberg, Phys. Rep. {\bf 463}, 1 (2008).

\bibitem{Szameit10}
A. Szameit and S. Nolte, J. Phys. B {\bf 43}, 163001 (2010).

\bibitem{Longhibook}
S. Longhi, {\it Control of photonic tunneling in coupled optical
waveguides}, in: Dynamical Tunneling: Theory and Experiment (Edited
by S. Keshavamurthy and  P. Schlagheck, CRC Press, Taylor \&
Francis, Boca Raton, FL, 2011), pp. 311-338.

\bibitem{B02}
U. Peschel, T. Pertsch, and F. Lederer, Opt. Lett. {\bf 23}, 1701;
R. Morandotti, U. Peschel, J. S. Aitchison, H. S. Eisenberg, and Y.
Silberberg, Phys. Rev. Lett. {\bf 83}, 4756 (1999); T. Pertsch, P.
Dannberg, W. Elflein, A. Br\"{a}uer, and F. Lederer, Phys. Rev.
Lett. {\bf 83}, 4752 (1999); G. Lenz, I. Talanina, and C.M. de
Sterke, Phys. Rev. Lett. {\bf 83}, 963 (1999); N. Chiodo, G. Della
Valle, R. Osellame, S. Longhi, G. Cerullo, R. Ramponi, P. Laporta,
and U. Morgner, Opt. Lett. {\bf 31}, 1651 (2006); H. Trompeter, W.
Krolikowski, D. N. Neshev, A. S. Desyatnikov, A.A. Sukhorukov, Yu.
S. Kivshar, T. Pertsch, U. Peschel, and F. Lederer, Phys. Rev. Lett.
{\bf 96}, 053903 (2006).

\bibitem{ZT}
R. Khomeriki and S. Ruffo, Phys. Rev. Lett. {\bf 94}, 113904 (2005);
H. Trompeter, T. Pertsch, F. Lederer, D. Michaelis, U. Streppel, A.
Br\"{a}uer, and U. Peschel, Phys. Rev. Lett. {\bf 96}, 023901
(2006); A. Fratalocchi, G. Assanto, K. A. Brzdakiewicz, and M. A.
Karpierz, Opt. Lett. {\bf 31}, 1489 (2006); A. Fratalocchi and G.
Assanto, Opt. Express {\bf 14}, 2021 (2006); S. Longhi, Europhys.
Lett. {\bf 76}, 416 (2006).

\bibitem{Dreisow09}
F. Dreisow, A. Szameit, M. Heinrich, T. Pertsch, S. Nolte, A.
T\"{u}nnermann, and S. Longhi, Phys. Rev. Lett. {\bf 102}, 076802
(2009).

\bibitem{DL}
S. Longhi, Opt. Lett. {\bf 30}, 2137 (2005); S. Longhi, M.
Marangoni, M. Lobino, R. Ramponi, P. Laporta, E. Cianci, and V.
Foglietti, Phys. Rev. Lett. {\bf 96}, 243901 (2006); R. Iyer, J. S.
Aitchison, J. Wan, M. M. Dignam, and C. M. de Sterke, Opt. Express
{\bf 15}, 3212 (2007); F. Dreisow, M. Heinrich, A. Szameit, S.
D\"{o}ring, S. Nolte, A. T\"{u}nnermann, S. Fahr, and F. Lederer,
Opt. Express {\bf 16}, 3474 (2008); A. Szameit, I.L. Garanovich, M.
Heinrich, A.A. Sukhorukov, F. Dreisow, T. Pertsch, S. Nolte, A.
T\"{u}nnermann, and Y.S. Kivshar, Nature Phys. {\bf 5}, 271 (2009);
A. Joushaghani, R. Iyer, J.K.S. Poon, J.S. Aitchison, C.M. de
Sterke, J. Wan, and M.M. Dignam, Phys. Rev. Lett. {\bf 103}, 143903
(2009); S. Longhi, Phys. Rev. B {\bf 80}, 235102 (2009); G. Della
Valle and S. Longhi, Opt. Lett. {\bf 35}, 673 (2010); A. Szameit,
I.L. Garanovich, M. Heinrich, A.A. Sukhorukov, F. Dreisow, T.
Pertsch, S. Nolte, A. Tunnermann, S. Longhi, and Y.S. Kivshar, Phys.
Rev. Lett. {\bf 104}, 223903 (2010).

\bibitem{CDT}
I. Vorobeichik, E. Narevicius, G. Rosenblum, M. Orenstein, and N.
Moiseyev, Phys. Rev. Lett. {\bf 90}, 176806 (2003); G. Della Valle,
M. Ornigotti, E. Cianci, V. Foglietti, and P. Laporta, and S.
Longhi, Phys. Rev. Lett. {\bf 98}, 263601 (2007); A. Szameit, Y. V.
Kartashov, F. Dreisow, M. Heinrich, T. Pertsch, S. Nolte, A.
T\"{u}nnermann, V. A. Vysloukh, F. Lederer, and L. Torner, Phys.
Rev. Lett. {\bf 102}, 153901 (2009).

\bibitem{stabi}
S. Longhi, D. Janner, M. Marano, and P. Laporta, Phys. Rev. E {\bf
67}, 036601 (2003); S. Longhi, M. Marangoni, D. Janner, R. Ramponi,
P. Laporta, E. Cianci, and V. Foglietti, Phys. Rev. Lett. {\bf 94},
073002 (2005).

\bibitem{AL}
T. Schwartz, G. Bartal, S. Fishman, and M. Segev, Nature {\bf 446},
55 (2007); Y. Lahini, A. Avidan, F. Pozzi, M. Sorel, R. Morandotti,
D. N. Christodoulides, and Y. Silberberg, Phys. Rev. Lett. {\bf
100}, 013906 (2008).

\bibitem{QZ}
S. Longhi, Phys. Rev. Lett. {\bf 97}, 110402 (2006); P. Biagioni, G.
Della Valle, M. Ornigotti, M. Finazzi, L. Du\'{o}, P. Laporta, and
S. Longhi, Opt. Express {\bf 16}, 3762 (2008); F. Dreisow, A.
Szameit, M. Heinrich, T. Pertsch, S. Nolte, A. T\"{u}nnermann, and
S. Longhi, Phys. Rev. Lett. {\bf 101}, 143602 (2008).

\bibitem{Rabi}
Y.V. Kartashov, V.A. Vysloukh and L. Torner, Phys. Rev. Lett. {\bf
99}, 233903 (2007); K. Shandarova, C.E. Ruter CE, D. Kip, K.G.
Makris, D.N. Christodoulides, O. Peleg, and M. Segev, Phys. Rev.
Lett. {\bf 102}, 123905 (2009);

\bibitem{STIRAP}
E. Paspalakis, Opt. Commun. {\bf 258}, 31 (2006); S. Longhi, G.
Della Valle, M. Ornigotti, and P. Laporta, Phys. Rev. B {\bf 76},
201101(R) (2007); Y. Lahini, F. Pozzi, M. Sorel, R. Morandotti, D.
N. Christodoulides, and Y. Silberberg, Phys. Rev. Lett. {\bf 101},
193901 (2008); F. Dreisow, A. Szameit, M. Heinrich, R. Keil, S.
Nolte, A. T\"{u}nnermann, and S. Longhi, Opt. Lett. {\bf 34}, 2405
(2009); F. Dreisow, M. Ornigotti, A. Szameit, M. Heinrich, R. Keil,
S. Nolte, A. Tunnermann, and S. Longhi Appl. Phys. Lett. {\bf 95},
261102 (2009).

\bibitem{molecule}
S. G. Krivoshlykov and I. N. Sissakian, Opt. Quantum Electron. {\bf
11}, 393 (1979); S. Longhi, Opt. Lett. {\bf 34}, 2736 (2009).

\bibitem{geometric}
A. Szameit, F. Dreisow, M. Heinrich, R. Keil, S. Nolte, A.
Tunnermann, and S. Longhi, Phys. Rev. Lett. {\bf 104}, 150403
(2010); G Della Valle and S Longhi, J. Phys. B {\bf 43}, 051002
(2010).

\bibitem{Kapitza}
S. Longhi, Opt. Lett. {\bf 36}, 819 (2011).

\bibitem{Longhi11}
S. Longhi, J. Phys. B {\bf 44}, 051001 (2011); S. Longhi, Phys. Rev.
A {\bf 83 },  034102 (2011); S. Longhi, Phys. Rev. A {\bf 83},
043835 (2011).

\bibitem{GR1}
K.S. Novoselov, A. K. Geim, S. V. Morozov, D. Jiang, M.I.
Katsnelson, I.V. Grigorieva, S.V. Dubonos, and A.A. Firsov, Nature
(London) {\bf 438}, 197 (2005); S.Y. Zhou, G.-H. Gweon, J. Graf,
A.V. Fedorov, C.D. Spataru, R.D. Diehl, Y. Kopelevich, D.-H. Lee,
Steven G. Louie, and A. Lanzara, Nature Phys. {\bf 2}, 595 (2006);
M.I. Katsnelson, K.S. Novoselov, and A.K. Geim, Nature Phys. {\bf
2}, 620 (2006).

\bibitem{GR2}
C. W. J. Beenakker, Rev. Mod. Phys. {\bf 80}, 1337 (2008); A. H.
Castro Neto, F. Guinea, N. M. Peres, K. S. Novoselov, and A. K.
Geim, Rev. Mod. Phys. {\bf 81}, 109 (2009).

\bibitem{ion}
P.M. Alsing, J.P. Dowling, and G.J. Milburn, Phys. Rev. Lett. 94,
220401 (2005); J. Schliemann, D. Loss, and R.M. Westervelt, Phys.
Rev. Lett. 94, 206801 (2005); A. Bermudez, M.A. Martin-Delgado, and
E. Solano, Phys. Rev. A {\bf 76},  041801(R) (2007); L. Lamata, J.
Leon, T. Schatz, and E. Solano, Phys. Rev. Lett. {\bf 98}, 253005
(2007); G. Juzeliunas, J. Ruseckas, M. Lindberg, L. Santos, and P.
Ohberg, Phys. Rev. A {\bf 77}, 011802(R) (2008); M. Johanning, A. F.
Var\'{o}n, and C. Wunderlich, J. Phys. B {\bf 42}, 154009 (2009); N.
Goldman, A. Kubasiak, A. Bermudez, P. Gaspard, M. Lewenstein, and
M.A. Martin-Delgado, Phys. Rev. Lett. {\bf 103}, 035301 (2009).

\bibitem{ion2}
S.L. Zhu, B.G. Wang, and L.M. Duan, Phys. Rev. Lett. {\bf 98},
260402 (2007); T.M. Rusin and W. Zawadzki, Phys. Rev. D {\bf 82},
125031 (2010); K.L. Wang, T. Liu, M. Feng, and K. Wang, Phys. Rev. A
{\bf 82}, 064501 (2010);  Q. Zhang, J.B. Gong, and C.H. Oh, Phys.
Rev. A {\bf 81}, 023608 (2010); D. Braun, Phys. Rev. A {\bf 82},
 013617 (2010); J. Casanova, J.J. Garcia-Ripoll, R. Gerritsma,
C.F. Roos, and E. Solano, Phys. Rev. A {\bf 82}, 020101 (2010); J.I.
Cirac, P. Maraner, and J.K. Pachos, Phys. Rev. Lett. {\bf 105},
190403  (2010).

\bibitem{Zitterbewegung}
K. Huang, Am. Phys. J. {\bf 20}, 479 (1952).

\bibitem{Klein}
O. Klein, Z. Phys. {\bf 53}, 157 (1929).

\bibitem{EGR1}
A. F. Young and P. Kim, Nat. Phys. {\bf 5}, 222 (2009); N. Stander,
B. Huard, and D. Goldhaber-Gordon, Phys. Rev. Lett. {\bf 102},
026807 (2009).

 \bibitem{EGR2}
G. A. Steele, G. Gotz and L. P. Kouwenhoven, Nature NanoTechn. {\bf
4}, 363 (2009).

\bibitem{KleinIons}
R. Gerritsma, B. P. Lanyon, G. Kirchmair, F. Z\"{a}hringer, C.
Hempel, J. Casanova, J. J. García-Ripoll, E. Solano, R. Blatt, and
C. F. Roos, Phys. Rev. Lett. {\bf 106}, 060503 (2011).

\bibitem{ZBion}
R. Gerritsma, G. Kirchmair, F. Z\"{a}hringer, E. Solano, R. Blatt,
and C. F. Roos, Nature (London) {\bf 463}, 68 (2010).

\bibitem{Haldane}
F.D.M. Haldane and S. Raghu, Phys. Rev. Lett. {\bf 100}, 013904
(2008); R.A. Sepkhanov, Ya. B. Bazaliy, and C.W.J. Beenakker, Phys.
Rev. A {\bf 75}, 063813 (2007); O. Peleg, G. Bartal, B. Freedman, O.
Manela, M. Segev, and D.N. Christodoulides, Phys. Rev. Lett. {\bf
98}, 103901 (2007); O. Bahat-Treidel, O. Peleg, and M. Segev, Opt.
Lett. {\bf 33}, 2251 (2008); T. Ochiai and M. Onoda, Phys. Rev. B
 {\bf 80}, 155103 (2009).

\bibitem{Zhang08}
X. Zhang, Phys. Rev. Lett. {\bf 100}, 113903 (2008).

\bibitem{Segev09}
O. Bahat-Treidel, O. Peleg, M. Grobman, N. Shapira, T. Pereg-Barnea,
and M. Segev, Phys. Rev. Lett. {\bf 104}, 063901 (2010); O.
Bahat-Treidel, O. Peleg, M. Segev, and H. Buljan, Phys. Rev. A {\bf
82}, 013830 (2010).

\bibitem{meta}
D.\"{O} G\"{u}ney and D.A. Meyer, Phys. Rev. A {\bf 79}, 063834
(2009); L.-G. Wang, Z.-G. Wang, J.-X. Zhang, and S.-Y. Zhu,
 Opt. Lett. {\bf 34}, 1510 (2009);  L.-G. Wang, Z.-G. Wang, and S.-Y. Zhu,
 EPL {\bf  86}, 47008 (2009).

\bibitem{Longhi09un}
S. Longhi, Opt. Lett. {\bf 35}, 235 (2010).

\bibitem{Longhi10Klein}
 S. Longhi, Phys. Rev. B {\bf 81}, 075102 (2010).
x
\bibitem{Longhi10Pair}
S. Longhi, Phys. Rev. A {\bf 81}, 022118 (2010).

\bibitem{DreisowPRL2010}
F. Dreisow, M. Heinrich, R. Keil, A. T\"{u}nnermann, S. Nolte, S.
Longhi, and A. Szameit, Phys. Rev. Lett. {\bf 105}, 143902 (2010).

\bibitem{BR1}
S. Longhi, Phys. Res. Int. {\bf 2010}, 645106 (2010).

\bibitem{BR2}
S. Longhi, Opt. Lett. {\bf 35}, 1302 (2010).

\bibitem{BR3}
S. Longhi, Cent. Eur. J. Phys. {\bf 9}, 110 (2011).

\bibitem{LonghiJPB2010}
S. Longhi, J. Phys. B {\bf 43}, 205402 (2010).

\bibitem{photonicgraphene}
S.H. Nam, J. Zhou, A.J. Taylor, and A. Efimov, Opt. Express {\bf
18}, 25329 (2010); M.I. Molina and Y.S. Kivshar, Opt. Lett. {\bf
35}, 2895 (2010); S. Bittner, B. Dietz, M. Miski-Oglu, P. Oria
Iriarte, A. Richter, and F. Sch\"{a}fer, Phys. Rev. B {\bf 82},
014301 (2010); M. Shen, L.X. Ruan, and X. Chen, Opt. Express {\bf
18}, 12779 (2010); S.H. Nam, A.J. Taylor, and A. Efimov, Opt.
Express {\bf 18}, 10120 (2010).

\bibitem{Longhi2010PRL}
S. Longhi, Phys. Rev. Lett. {\bf 105}, 013903 (2010).

\bibitem{SzameitTachions}
A. Szameit, M.C. Rechtsman, O. Bahat-Treidel, and M. Segev, "Complex
photonic graphene: Optical tachyons, strain, and PT-symmetry ",
arXiv:1103.3389 (2011).

\bibitem{ZBS}
E. Schr\"{o}dinger, Sitz. Preuss. Akad. Wiss. Phys.-Math. Kl. {\bf
24}, 418 (1930).

\bibitem{Grenier}
W. Greiner, {\it Relativistic Quantum Mechanics} (Berlin, Springer,
1990).

\bibitem{Cannata90}
F. Cannata, L. Ferrari, and G. Russo, Solid State Commun. {\bf 74},
309 (1990).

\bibitem{Calogeracos99}
A. Calogeracos and N. Dombey, Contemp. Phys. {\bf 40}, 313 (1999).

\bibitem{Sauter}
F. Sauter, Z. Phys. {\bf 69}, 742 (1931).

\bibitem{Emilio}
P. Christillin and E. d'Emilio, Phys. Rev. A {\bf 76}, 042104
(2007).
%

\bibitem{libro1}
E. S. Fradkin, D. M. Gitman, and Sh. M. Shvartsman, {\it Quantum
Electrodynamics with Unstable Vacuum} (Springer, Berlin, 1991).

\bibitem{libro2}
H. K. Avetissian, {it Relativistic Nonlinear Electrodynamics}
(Springer, New York, 2006).

\bibitem{Swinger}
J. Schwinger, Phys. Rev. {\bf 82}, 664 (1951).

\bibitem{Brezin}
E. Brezin and C. Itzykson, Phys. Rev. D {\bf 2}, 1191 (1970).

\bibitem{32}
M. Ruf, G.R. Mocken, C. M\"{u}ller, K.Z. Hatsagortsyan, and C.H.
Keitel, Phys. Rev. Lett. {\bf 102}, 080402 (2009).

\bibitem{Dir1}
M. Moshinsky and A. Szczepaniak, J. Phys. A {\bf 22}, L817 (1989).

\bibitem{Dir2}
M. Moshinsky and Y.F. Smirnov, {\it {The Harmonic Oscillator in
Modern Physics}} (Harwood, Amsterdam, 1996).

\bibitem{Dir3}
 R.P. Martinez-y-Romero, H.N. Nunez-Yepez, and A.L. Salas-Brito,
 Eur. J. Phys. {\bf 16}, 135 (1995).

\bibitem{Dir4}
D. Ito, K. Mori, and E. Carrieri, Nuovo Cimento {\bf 51}A, 1119
(1967).

\bibitem{Dir5}
J. Bentez, R.P. Martnez-y-Romero, H.N. Nuez-Yepez, and A.L.
Salas-Brito, Phys. Rev. Lett. {\bf 64}, 1643 (1990); see also: {\it
Erratum}, Phys. Rev. Lett. {\bf 65}, 2085 (1990).

\bibitem{Dir6}
R.P. Martriaanez-y-Romero, M. Moreno, and A. Zentella, Phys. Rev. D
{\bf 43},  2036 (1991).

\bibitem{Dir7}
F. Dominguez-Adame and M. A. Gonzalez , Europhys. Lett. {\bf 13},
193 (1990).


\bibitem{KP}
R. de L. Kronig and W.G. Penney, Proc. Roy. Soc. London Ser. A {\bf
130}, 499 (1931).

\bibitem{RKP}
F. Dominguez-Adame, Am. J. Phys. {\bf 55}, 1003 (1987).

\bibitem{Dr0}
S.G. Davison and M. Streslicka, J. Phys. C {\bf 2}, 1802 (1969).

\bibitem{Dr1}
B. H. J. McKellar and G.J. Stephenson, Phys. Rev. A {\bf 36}, 2566
(1987).

\bibitem{Dr2}
F. Dominguez-Adame, J. Phys.: Condes. Matter {\bf 1}, 109 (1989).

\bibitem{Dr3}
I.M. Mladenov, Phys. Lett. A {\bf 131}, 313 (1989).

\bibitem{Dr4}
G.J. Clerck and B.H.J. McKellar, Phys. Rev. C {\bf 41}, 1198 (1990).

\bibitem{Dr5}
C.L. Roy and C. Basu, J. Phys. Chem. Solids {\bf 52}, 745 (1991).

\bibitem{Dr6}
F. Dominguez-Adame and A. Sanchez, Phys. Lett. A {\bf 159}, 153
(1991).

\bibitem{Dr7}
G.J. Clerck and B.H.J. McKellar, Phys. Rev. B {\bf 47}, 6942 (1993).

\bibitem{Dr8}
C. Basu, C.L. Roy, E. Macia, F. Dominguez-Adame, and A. Sanchez, J.
Phys. A {\bf 27}, 3285 (1994).

\bibitem{Dr9}
F. Dominguez-Adame, E. Macia, A. Khan, and C.L. Roy, Physica B {\bf
212}, 67 (1995).

\bibitem{Dr10}
M. Barbier, F.M. Peeters, P. Vasilopoulos, and J.M. Pereira, Phys.
Rev. B {\bf 77}, 115446 (2008).

\bibitem{SS1}
M.L. Glasser and S.G. Davison, Int. J. Quantum Chem. {\bf 4}, 867
(1970).

\bibitem{SS2}
S.G. Davison and J.D. Levine, Solid State Phys. {\bf 25}, 32 (1970).

\bibitem{SS3}
M. Steslicka and S.G. Davison, Phys. Rev. B {\bf 1},  1858 (1970).

\bibitem{SS4}
R Subramanian and K.V. Bhagwat, J. Phys. C {\bf 5}, 798 (1972).

\bibitem{SS5}
S. Yuyi, Surface Science {\bf 108}, L477 (1981).

\bibitem{SS6}
C.L. Roy and J.S. Pandey, Physica {\bf 137A}, 389 (1986).

\bibitem{Longhi1}
D. Janner, G. Galzerano, G. Della Valle, P. Laporta, S. Longhi, and
M. Belmonte, Phys. Rev. E {\bf 72}, 056605 (2005); S. Longhi, D.
Janner, G. Galzerano, G. Della Valle, D. Gatti, and P. Laporta,
Electron. Lett. {\bf 41}, 1075 (2005).

\bibitem{Bender98}
C.M. Bender and S. Boettcher, Phys. Rev. Lett. {\bf 80}, 5243 (1998)

\bibitem{Bender02}
C. M. Bender, D.C. Brody, and H. F. Jones, Phys. Rev. Lett. {\bf
89}, 270401 (2002).

\bibitem{Mostafazadeh02}
A. Mostafazadeh, J. Math. Phys. {\bf 43}, 2814 (2002).

\bibitem{BenderReview}
C.M. Bender, Rep. Prog. Phys. {\bf 70}, 947 (2007).

\bibitem{MostafazadehReview}
A. Mostafazadeh, e-print arXiv:0810.5643.

\bibitem{RotterReview}
N. Moiseyev, Phys. Rep. {\bf 302}, 211 (1998); J.G. Muga, J.P.
Palao, B. Navarro, and I.L. Egusquiza, Phys. Rep. {\bf 395}, 357
(1998); I. Rotter, J. Phys. A {\bf 42}, 1 (2009).

\bibitem{O1}
A. Ruschhaupt, F. Delgado, and J. G. Muga, J. Phys. A {\bf 38}, L171
(2005).

\bibitem{O2}
R. El-Ganainy, K. G. Makris, D. N. Christodoulides, and Z. H.
Musslimani, Opt. Lett. {\bf 32}, 2632 (2007); K.G. Makris, R.
El-Ganainy, D.N. Christodoulides, and Z.H. Musslimani, Phys. Rev.
Lett. {\bf 100}, 103904 (2008).

\bibitem{O3}
A. Guo, G.J. Salamo, D. Duchesne, R. Morandotti, M. Volatier-Ravat,
V. Aimez, G. A. Siviloglou, and D. N. Christodoulides, Phys. Rev.
Lett. {\bf 103}, 093902 (2009).

\bibitem{O4}
 S. Klaiman, U. G\"{u}nther, and N.
Moiseyev, Phys. Rev. Lett. {\bf 101}, 080402 (2008).

\bibitem{O5}
A. Mostafazadeh, Phys. Rev. Lett. {\bf 102}, 220402 (2009).

\bibitem{O6}
S. Longhi, Phys. Rev. Lett. {\bf 103}, 123601 (2009); S. Longhi,
Phys. Rev. B {\bf 80}, 235102 (2009); S. Longhi, Phys. Rev. A {\bf
81}, 022102 (2010).

\bibitem{O7}
C.E. R\"{u}ter, K.G. Makris, R. El-Ganainy, D.N. Christodoulides, M.
Segev, and D. Kip, Nature Phys. {\bf 6}, 192 (2010).

\bibitem{D1}
C. Mudry, B.D. Simons, and A. Altland, Phys. Rev. Lett. {\bf 80},
4257 (1998); H. Egrifes and R. Sever, Phys. Lett. A {\bf 344}, 117
(2005); A. Sinha and P. Roy, Mod. Phys. Lett. A {\bf 20}, 2377
(2005); C.S Jia and A. de Souza Dutra, J. Phys. A {\bf 39}, 11877
(2006); F. Cannata and A. Ventura, Phys. Lett. A {\bf 372}, 941
(2008); O. Mustafa, S.H. Mazharimousavi, Int. J. Theor. Phys. {\bf
47}, 1112 (2008).

\bibitem{D2}
C.-S. Jia and A. de Souza Dutra, Ann. Phys. {\bf 323}, 566 (2008);
C-S. Jia, P.-Q. Wang, J.-Yi Liu, and S. He, Int. J. Theor. Phys.
{\bf 47}, 2513 (2008).

\bibitem{D3}
 V.G.C.S. dos Santos, A. de Souza Dutra, and
M.B. Hott, Phys. Lett. A {\bf 373}, 3401 (2009); F. Cannata and A.
Ventura, J. Phys. A {\bf 43}, 075305 (2010).

\bibitem{field}
C.M. Bender, K.A. Milton, and Van M. Savage, Phys. Rev. D {\bf 62},
085001 (2000); C.M. Bender, D.C. Brody, and H.F. Jones, Phys. Rev.
Lett. {\bf 93}, 251601  (2004); C.M. Bender, S.F. Brandt, J.-H.
Chen, and Q. Wang, Phys. Rev. D {\bf 71}, 065010 (2005); A.
Mostafazadeh, Int. J. Mod. Phys. A {\bf 21}, 2553 (2006).

\bibitem{Poladian}
J. Carroll, J. Whiteaway, and D. Plumb, {\it Distributed feedback
semiconductor lasers} (The Institution of Electrical Engineers,
London, 1998).

\bibitem{Longhi2010PRA}
S. Longhi, Phys. Rev. A {\bf 82}, 031801(R) (2010); Y.D. Chong, Li
Ge, and A.D. Stone, Phys. Rev. Lett. {\bf 106}, 093902 (2011).

\bibitem{tachy1}
A. Chodosa and A.I. Hausera, Phys. Lett. B {\bf 150}, 431 (1985); J.
Ciborowski1 and  J. Rembielinski, Eur. Phys. J. C {\bf 8}, 157
(1999); T. Chang and G. Ni, Fizika B {\bf 11}, 49 (2002).

\bibitem{tachy2}
G. Feinberg, Phys. Rev. Lett. {\bf 159}, 1089 (1967); O.M.P.
Bilaniuk,and E.C.G. Sudarshan, Physics Today {\bf 22}, 43 (1969).

\bibitem{LonghiOL10}
S. Longhi, Opt. Lett. {\bf 35}, 3844 (2010).

\end{thebibliography}
\end{document}